\documentclass[12pt,preprint]{aastex}

\newcommand{\iras}{IRAS~00420$+$5530}
\newcommand{\prcala}{J0042$+$5708}
\newcommand{\prcalb}{J0047$+$5657}
\newcommand{\kmps}{km s$^{-1}$}
\newcommand{\To}{\Theta_{\circ}}
\newcommand{\Ro}{R_{\circ}}
\newcommand{\water}{H$_2$O}
\newcommand{\etal}{et al.}

\slugcomment{To appear in the {\em Astrophysical Journal, March 2009}}

\shorttitle{Parallax Distance to \iras}
\shortauthors{Moellenbrock et al.}

\begin{document}

\title{A Precise Distance to \iras\ via \\
  \water\ Maser Parallax with the VLBA}

\author{G. A. Moellenbrock, M. J. Claussen, and W. M. Goss}
\affil{National Radio Astronomy Observatory, Socorro, NM 87801}
\email{gmoellen@nrao.edu}

\begin{abstract}
We have used the VLBA to measure the annual parallax of the \water\ masers
in the star-forming region \iras.  This measurement yields a direct
distance estimate of $2.17\pm 0.05$ kpc ($<$3\%), which disagrees
substantially with the standard kinematic distance estimate of
$\sim$4.6 kpc \citep[according to the rotation curve of][]{Brand93}, as
well as most of the broad range of distances (1.7$-$7.7 kpc) used in
various astrophysical analyses in the literature.  The 3-dimensional
space velocity of \iras\ at this new, more accurate distance implies a
substantial non-circular and anomalously slow Galactic orbit,
consistent with similar observations of W3(OH)
\citep{Xu06,Hachisuka06}, as well as line-of-sight velocity residuals
in the rotation curve analysis of \citet{Brand93}.  The Perseus spiral
arm of the Galaxy is thus more than a factor of two closer than
previously presumed, and exhibits motions substantially at odds
with axisymmetric models of the rotating Galaxy.

\end{abstract}



\section{Introduction}

Distance estimates to celestial objects are one of the most
fundamental measurements in astronomy and astrophysics.  Knowledge of
the distance to astronomical sources in the sky are needed in order to
estimate their physical properties, such as luminosities, masses,
kinematics, and dynamics.  In recent years, the technique of Very Long
Baseline Interferometry (VLBI) has been used to make precise
astrometric measurements (to a precision of of a few tens of
microarcseconds for a single observation) and thus holds the promise
to extend the range of the direct distance measurements of annual
parallax up to at least 10 kpc with 10\% accuracy using radio
telescopes such as the Very Long Baseline Array (VLBA) of the
National Radio Astronomy Obsevatory (NRAO).

Compact, bright radio-emitting objects such as pulsars and masers are
choice beacons for VLBI distance measurements across the Galaxy
\citep[e.g.,][]{vLangevelde00,Brisken02,Chatterjee04,Xu06}.  Such
distance measurements enable enormous improvements in our
understanding of Galactic structure and also the physics of individual
objects.  \citet{Reid08} reviewed the importance of
measuring parallaxes and proper motions to help delineate Galactic
structure and showed, in a survey for parallaxes and proper motions of
young, high-mass stars, that kinematic distances are systematically
too large.

In this paper, we report on using the VLBA to measure the annual
parallax of the star-forming region \iras, using the bright water
masers associated with the IRAS source.  \iras\ $(l=122.0^\circ$,
$b=-7.1^\circ)$ is a star formation region exhibiting a molecular
outflow \citep{Zhang05}, 3.6cm and 3mm continuum emission
\citep{Molinari02}, and \water\ masers \cite[e.g.,][]{Brand94}.  It 
is also coincident with dense gas traced by HCO$^+$(1-0), indicating
that the cluster of young stars remains deeply embedded in its natal
cloud of molecular gas \citep{Molinari02}.  The systemic velocity of
the gas around \iras\ is $\sim-$51 \kmps\ with respect to the
local standard of rest (LSR) based on the measurements of
ammonia radial velocities \citep{Molinari96} and the HCO$^+$(1-0)
\citep{Molinari02}.  The water masers occur at similar line-of-sight
velocities as the surrounding gas, but with a somewhat broader range 
(LSR velocities $-$52 to $-$40 \kmps).

Distances to \iras\ of 4.3 to 7.7 kpc (presumably kinematic, in most
cases) have been used in recent papers, despite a much closer
photometric distance of 1.7 kpc \citep{Neckel84}.  Reflecting this
broad range of distance estimates, luminosities of the IRAS source
(presumably the exciting star) between $1.2\times10^4$ and
$5.2\times10^4 L_{\odot}$ have been reported
\citep[e.g.,][]{Molinari02,Zhang05}.  Clearly, a better distance
estimate is desirable for this object.  In addition, \iras\ lies in
the direction of the Perseus arm, which has been the topic of a recent
distance study by \citet{Xu06}.  \citet{Xu06} addressed the large
discrepancy between the luminosity and kinematic distance estimates in
the Perseus arm by accurately measuring the distance (using VLBI
techniques) of the methanol masers in the compact HII region W3(OH).
It is important that distances to other objects in the direction of
the Perseus arm be precisely measured to compare to the work presented
by \citet{Xu06}, and to determine small-scale deviations from the
distances and kinematics in this region which could be due to peculiar
motions of the studied objects.

We present here a careful analysis of the observing style, calibration
techniques, data reduction and parallax/proper motion fitting in order
to provide the error analysis required to derive the most precise
measurements of the distance to \iras.

\section{Observations}

Between 2005 November and 2006 September, we made twelve 4-hour VLBA
observations (approximately monthly) of the \water\ masers in
\iras\ to measure their annual parallax and proper motion.
The large number of monthly epochs were chosen to ensure that the
parallax and proper motion fits would not be confused by maser spots
fading and disappearing between epochs. In the end, most of the spots
persisted through most epochs, and this was not a problem.  The
observations were made from the VLBA's dynamic observing queue,
wherein our standardized observing script was initiated opportunistically
and periodically in real-time according to availability of the array, 
scientific ranking (relative to fixed-schedule programs and other 
dynamically scheduled observations) and other operational constraints, 
such as observing direction and weather. Since \iras\ is at 
relatively high declination (and thus circumpolar at most VLBA sites), 
our four-hour observation tended to rise to the top of the dynamic queue 
quite easily and often as a convenient schedule-filler, and so the interval 
between epochs was often shorter than one month.  The properties of
the observations are listed in Table~\ref{tab1}.

\begin{deluxetable}{cllll}
\tabletypesize{\scriptsize}
\rotate
\tablecaption{Observations\label{tab1}}
\tablewidth{0pt}
\tablehead{
\colhead{Epoch} & \colhead{Date} & \colhead{Time range} & \colhead{Beam} & \colhead{Note} \\
\colhead{} & \colhead{} & \colhead{(UTC)} & \colhead{} & \colhead{}
}
\startdata
A & 2005 Nov 24 & 02:30--06:30 & $1.00\times0.34$ @ -17.5$^\circ$ & Brewster did not participate \\
B & 2005 Dec 19 & 00:51--04:51 & $0.98\times0.33$ @ -16.6$^\circ$ & \\
C & 2006 Jan 08 & 23:33--03:33 & $0.91\times0.37$ @ -16.3$^\circ$ & \\
D & 2006 Jan 26 & 22:22--02:22 & $0.86\times0.46$ @ -23.7$^\circ$ & \\
E & 2006 Feb 22 & 20:32--00:32 & $0.88\times0.36$ @ -18.3$^\circ$ & \\
F & 2006 Mar 17 & 19:01--23:01 & $0.87\times0.37$ @ -19.3$^\circ$ & \\
G & 2006 Apr 08 & 17:35--21:35 & $0.88\times0.37$ @ -14.3$^\circ$ & \\
H & 2006 May 08 & 15:37--19:37 & $0.84\times0.62$ @ -24.9$^\circ$ & \\
I & 2006 May 31 & 13:51--17:51 & $0.85\times0.36$ @ -13.3$^\circ$ & \\
J & 2006 Jun 22 & 12:40--16:40 & $0.85\times0.36$ @ -15.3$^\circ$ & \\
K & 2006 Aug 04 & 09:51--13:51 & $0.93\times0.76$ @ \phm{-}21.7$^\circ$ & Mauna Kea did not participate; epoch not used \\
L & 2006 Sep 02 & 07:57--11:57 & $0.84\times0.34$ @ -16.7$^\circ$ & \\
\enddata
\end{deluxetable}

Four 16 MHz spectral windows were simultaneously observed in the
VLBA's 22 GHz band.  One spectral window (the second) was set up to
cover the \water\ maser lines in \iras\ at $V_{LSR}\sim-$46 \kmps.
The other three windows were distributed across the 500 MHz instantaneous
IF bandwidth of the VLBA to provide optimal delay sensitivity for
calibration of the troposphere.  

\begin{figure}
\includegraphics[angle=-90,scale=0.70]{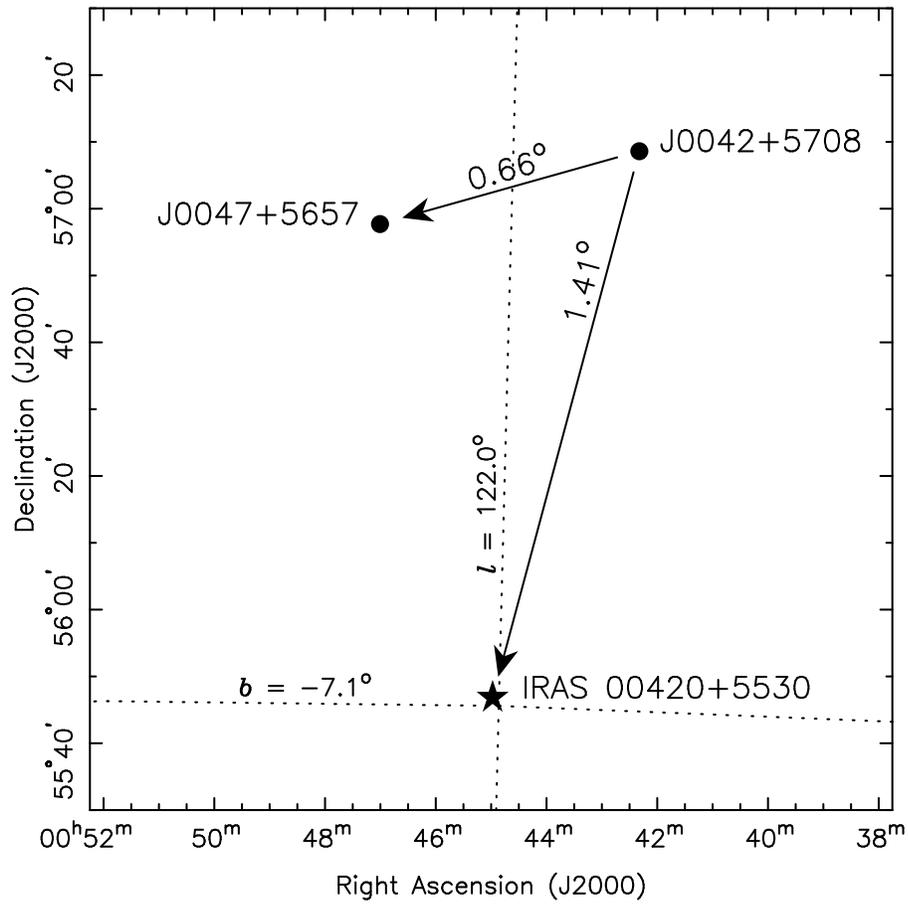}
\caption{Phase referencing geometry.  \prcala\ is the primary phase-reference calibrator; its
phase solutions are used to calibrate both \iras\ and
\prcalb. \label{calgeo}}
\end{figure}

In addition to \iras, nine compact background radio-loud quasars were
observed as calibrators.  Two of these calibrators were for
phase-referencing purposes to provide for accurate relative astrometry
(see Table~\ref{tab2} and figure~\ref{calgeo}).  The first of these
calibrators (\prcala, $\sim$1.41$^\circ$ from \iras) was observed for
primary phase referencing.  For most of the observation, \iras\ and
\prcala\ were observed alternately on a $\sim$60s timescale (30s dwell on
each).  Once per hour, an alternate calibrator (\prcalb,
$\sim$0.66$^\circ$ from \prcala) was observed to permit verification of
the phase-referencing calibration.


\begin{deluxetable}{lllll}
\tabletypesize{\scriptsize}
\rotate
\tablecaption{Source Data\label{tab2}}
\tablewidth{0pt}
\tablehead{
\colhead{Source Name} & \colhead{R.A.} & \colhead{Decl.} & \colhead{Flux Density} & \colhead{Note} \\
\colhead{} & \colhead{(J2000.0)} & \colhead{(J2000.0)} & \colhead{(Jy)} & \colhead{}
}
\startdata
\iras    & 00$^h$44$^m$58$^s$.397  & 55$^\circ$46'47''.600 & up to $\sim$70.0 & Science Target: \water\ masers \\
\prcala  & 00$^h$42$^m$19$^s$.4517 & 57$^\circ$08'36''.586 & $\sim$0.18       & Primary phase-reference calibrator \\
\prcalb  & 00$^h$47$^m$00$^s$.4288 & 56$^\circ$57'42''.395 & $\sim$0.10       & Secondary phase-reference calibrator \\
\enddata


\end{deluxetable}

The other seven calibrators, chosen from the VLBA calibrator
list \citep{Ma98} were observed as a group once per
hour (five times in four hours) to enable determination of the
residual macroscopic troposphere delay error at each antenna.

The accuracy of phase-referencing calibration at the VLBA is limited
by an zenith angle-dependent delay error arising from inaccuracies in
the macroscopic troposphere model used at the VLBA correlator
\citep{Reid99}.  The seasonal zenith troposphere delay model is typically
in error by many centimeters, and this leads to a significant
time-dependent delay error between the science target and
phase-reference calibrator since they are observed at slightly
different zenith angles.  As a result, the phase-referencing
determined for the calibrator does not fully calibrate the target
unless this macroscopic component is compensated.

This effect is calibrated by observing a set of bright ($\gtrsim 1$
Jy) calibrators at a wide range of zenith angles and fitting the
observed residual delays for the zenith troposphere error, assuming
the delay increases away from the zenith as $\sec z$, where $z$ is the
zenith angle.  The standard approach \citep[e.g.,][]{Reid04} is
to select approximately one dozen such calibrators distributed over
the entire sky and observe them 1$-$3 times over the course of the
observation to get a few instantaneous measurements of the zenith
delay.  Such an approach does not permit tracking the delay error in
time very accurately and makes no allowance for any azimuth dependence
of the delay error, such as might be expected as weather changes over
an antenna.  Our approach was to select a smaller set of calibrators
distributed sufficiently in zenith angle ($1 < \sec z < 2.5$) at the
approximate azimuth of our target, and observe them more continuously
during the observation.

After each observation, the data were correlated in two passes.  The
first pass correlated all four spectral windows at low spectral
resolution (32 channels).  The second pass correlated only the
spectral window containing the \water\ maser lines with 1024 channels,
yielding a channel spacing of 15.625 kHz, and thus a spectral resolution
in the LSR of 18.75 kHz, or 0.253 \kmps.

\section{Data Reduction}

The visibility calibration was performed in NRAO AIPS
\citep{Greisen03}, following the exact same procedure for each epoch.
The low-spectral-resolution data were processed first, as follows.

First, several standard a priori calibrations were applied, including
revised Earth orientation parameter corrections (EOP, important for
good astrometry), parallactic angle phase, quantized sampling bias
amplitude corrections, and system temperature/gain amplitude
calibration.  

Second, a scan on one of the bright macroscopic troposphere
calibrators (J2202$+$4216$=$BL~Lac, $\sim$3.0 Jy) was selected to
serve as a reference for all subsequent phase and delay calibration
solutions.  For this scan, the net residual delay and non-linear
bandpass was determined in each spectral window.  This solution,
applied to the rest of the data, removes constant instrumental phase
and delay errors, introduces a tropospheric delay residual offset
(whatever the troposphere delay was toward J2202$+$4216 in this scan),
and leaves behind only relative (and relatively small) delay residuals
that are time- and direction-dependent.  These errors include any
residual instrumental clock drift, zenith-angle-dependent relative
macroscopic troposphere delay errors (to be solved using the seven
bright calibrators), and time-dependent (phase-) delay errors (to be
phase-referenced).

The macroscopic troposphere delay residual calibration is obtained by
applying the calibrations described above and solving for the phase in
each of the four spectral windows and in each of the scans on the
seven bright calibrators.  
These phases are then fit for multi-band delays.  In general, the
likelihood of ambiguity resolution problems in the multi-band delay
determination are small, since these delays are residuals, and the
expected zenith delay errors are expected to be only a fraction of a
nanosecond.  The multi-band delays are then used to jointly solve for
the residual instrumental clock drift and time-dependent zenith
troposphere delay error, assuming the $\sec z$ model.  The separation
of these two effects depends upon observing the calibrators out of
zenith angle order, else a monotonic clock drift would be difficult to
distinguish from a component of the elevation-dependent troposphere
delay residual.

Solutions with constant, linear, quadratic, and cubic time-dependence
(and all combinations) in both clock and troposphere delay terms were
attempted.  The higher-order fits tended to diverge. Optimal fits were
obtained for quadratic troposphere and linear clock.  Typical zenith
delay residuals were found to be a few to 15cm, with variations over
four hours at the 1$-$2 cm level.  The overall accuracy of this delay
calibration is approximately 0.5$-$1.0 cm, and is limited by the
simplicity of the zenith angle dependence and unmodeled local
time-variation of the troposphere delay (of the sort we would
otherwise phase-reference). Since this is a residual effect, negative
troposphere delay solutions are possible.  Also, the troposphere delay
solution is not referenced to any specific antenna since the
zenith angle dependence and differential zenith angle sampling at each
antenna breaks the usual degeneracy.

\begin{figure}
\includegraphics[angle=-90,scale=0.7]{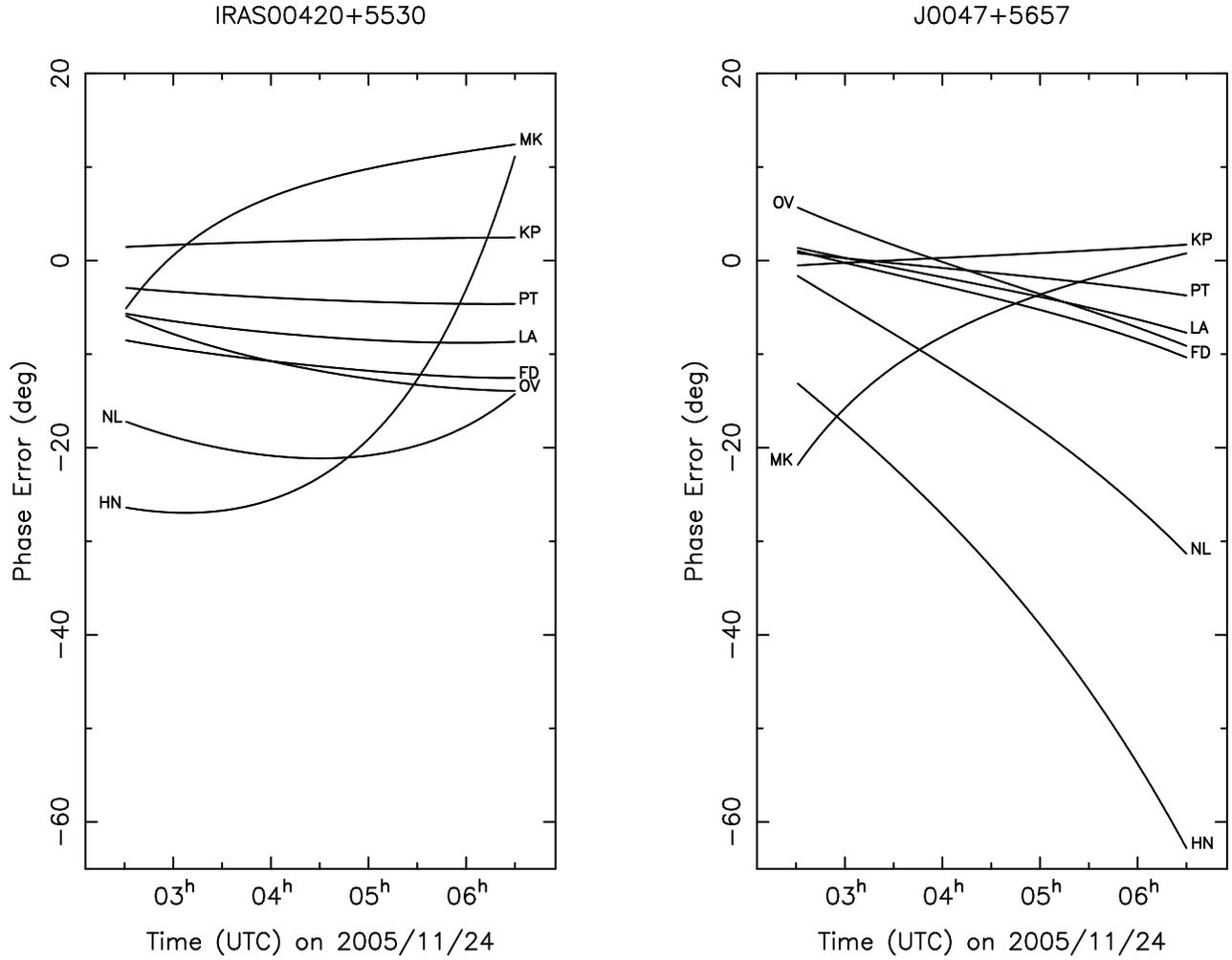}
\caption{Time dependent phase errors due to macroscopic troposphere
delay residuals (correlator zenith delay model), per antenna. Right: for
\iras.  Left: for the secondary calibrator, \prcalb. \label{mtpherr}}
\end{figure}

Figure \ref{mtpherr} shows, for epoch A, the systematic
phase-referencing errors for \iras\ and \prcalb\ due to their
time-dependent zenith angle difference relative to \prcala\ and the (mean)
zenith delay model calibration determined as described above.  The
zenith delay errors range in magnitude from 0.6 cm for KP to 13.5 cm
for HN (BR did not participate, and SC was excluded).  The errors are
generally larger for \iras\ since it has a larger separation from
\prcala, and at a relatively unfavorable position angle (near transit)
that maximizes the zenith angle difference.  Only for the easternmost
antennas in the array (HN, NL), which are observing farthest from
transit, is the position angle of \prcalb\ relative to \prcala\
sufficient to excessively amplify this systematic phase error.

The VLBA station at St. Croix (SC) was generally the worst performing
station in the delay calibration, yielding poor multi-band delay fits
and/or poor zenith delay solutions.  This is caused by excessive and
highly variable tropospheric water vapor at SC's sea-level location
and humid climate.  Since this station also yields the least accurate
phase referencing solutions for the same reasons, it was excised from
all epochs.  This results in the loss of many relatively long
baselines, but the poor calibration of these baselines would be of
questionable value.

At this point, the high-spectral-resolution data were processed, first
with all of the same {\em a priori} calibrations as the low-resolution data.
The tropospheric delay calibration solution (including the clock
drift) was then transferred from the low-resolution dataset, to be
used as an additional a priori calibration for the high-resolution
data.  The same strong calibrator scan was selected to obtain the
global instrumental reference calibration, except a polynomial
bandpass was used since the per-channel SNR was insufficient for a
sampled bandpass at the high spectral resolution.  Finally, the
phase-reference calibration was determined on \prcala.  The full
calibration was then applied to \iras, \prcala, and the secondary
phase-reference calibrator, \prcalb.  After re-sampling the velocity
axis for each epoch to the LSR, the data were exported in UVFITS
format.

\begin{figure}
\includegraphics[angle=-90,scale=1.0]{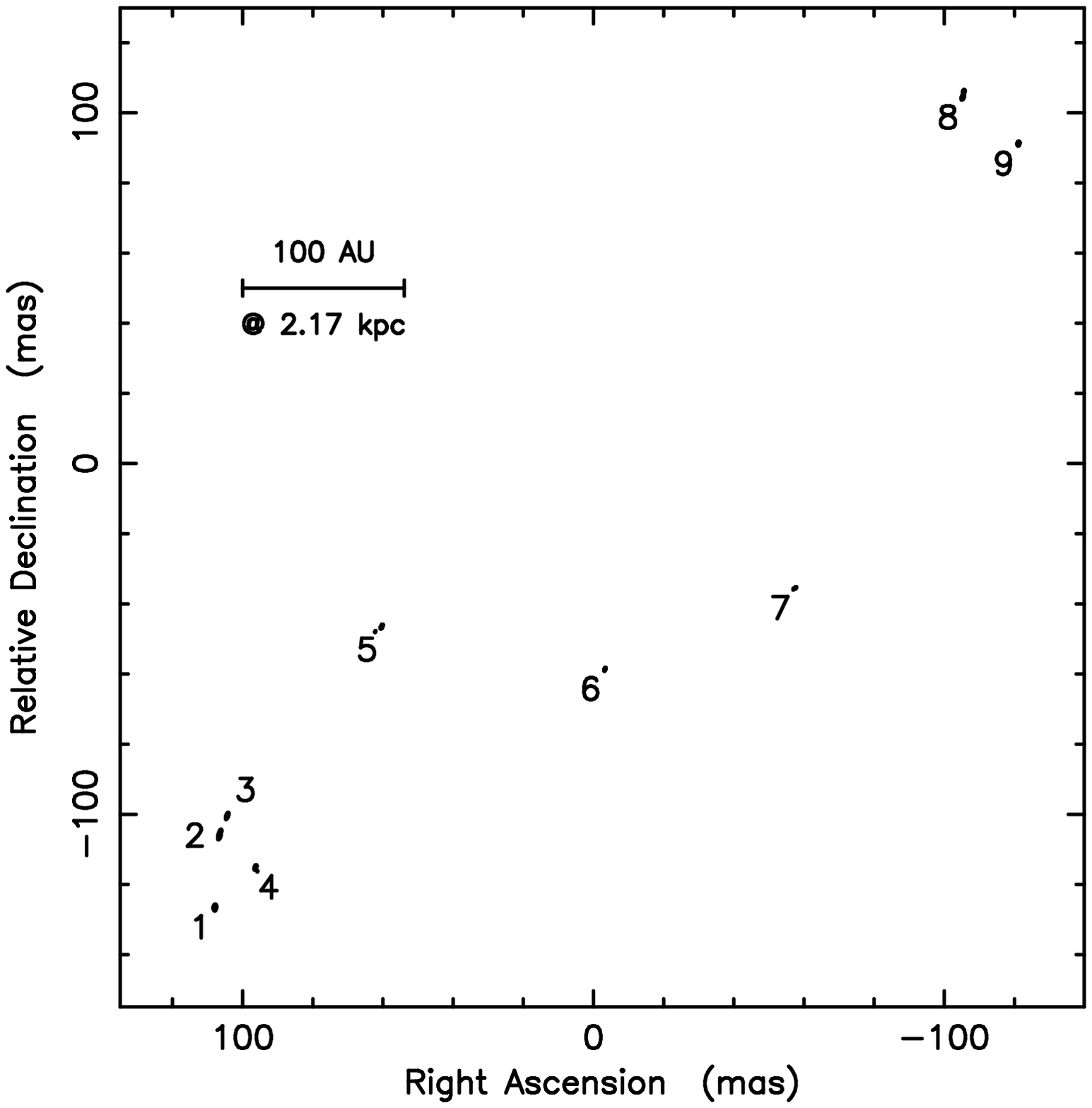}
\caption{Spatial distribution of maser regions (epoch A).  The phase
center of the observation is at (100, -100).  This image has a
resolution of 1.01 $\times$ 0.354 milliarcseconds at position angle
$-19^\circ$, a peak flux density of 74.5 Jy beam$^{-1}$ and a minimum
contour of 10 Jy beam$^{-1}$.  Each region is numbered for reference
in the text and tables. The linear scale corresponding to the derived
distance (described in the text) is indicated.
\label{skyfig}}
\end{figure}

The data for each epoch were then imaged in Difmap \citep{Shepherd97},
selecting the channels with maser emission.  First, a relatively low
spatial resolution image of a wide field-of-view was generated to
spatially locate significant maser regions, then each of these regions
were imaged at full resolution.  A simple multi-field clean
deconvolution script was developed in Difmap to optimize the imaging
of channels with signal in more than one region.  Figure \ref{skyfig}
shows the spatial distribution of the maser regions (integrated clean
components from epoch A), integrated in velocity.  There are nine
major maser spot regions, two of which (4 and 7) contain two distinct
maser spots, with significant emission in a total of 44 channels.  The
typical imaging sensitivity achieved per channel was 100$-$200
mJy/beam, with maser peaks ranging up to $\sim$75 Jy beam$^{-1}$.

\begin{figure}
\includegraphics[angle=0,scale=0.8]{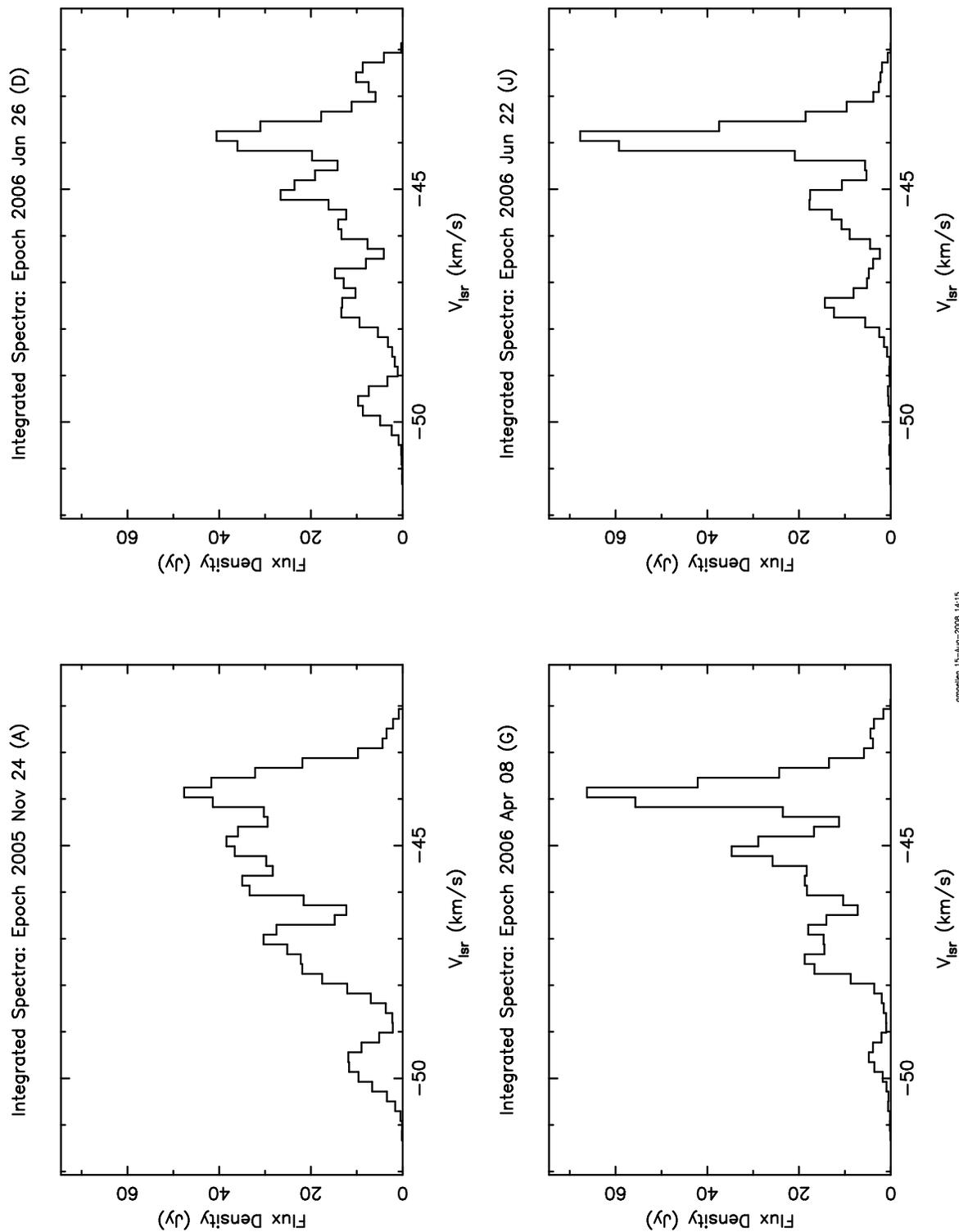}
\caption{Integrated spectra of \iras\ at epochs A, D, G, and J. \label{integspecfig}}
\end{figure}

\begin{figure}
\includegraphics[angle=0,scale=0.8]{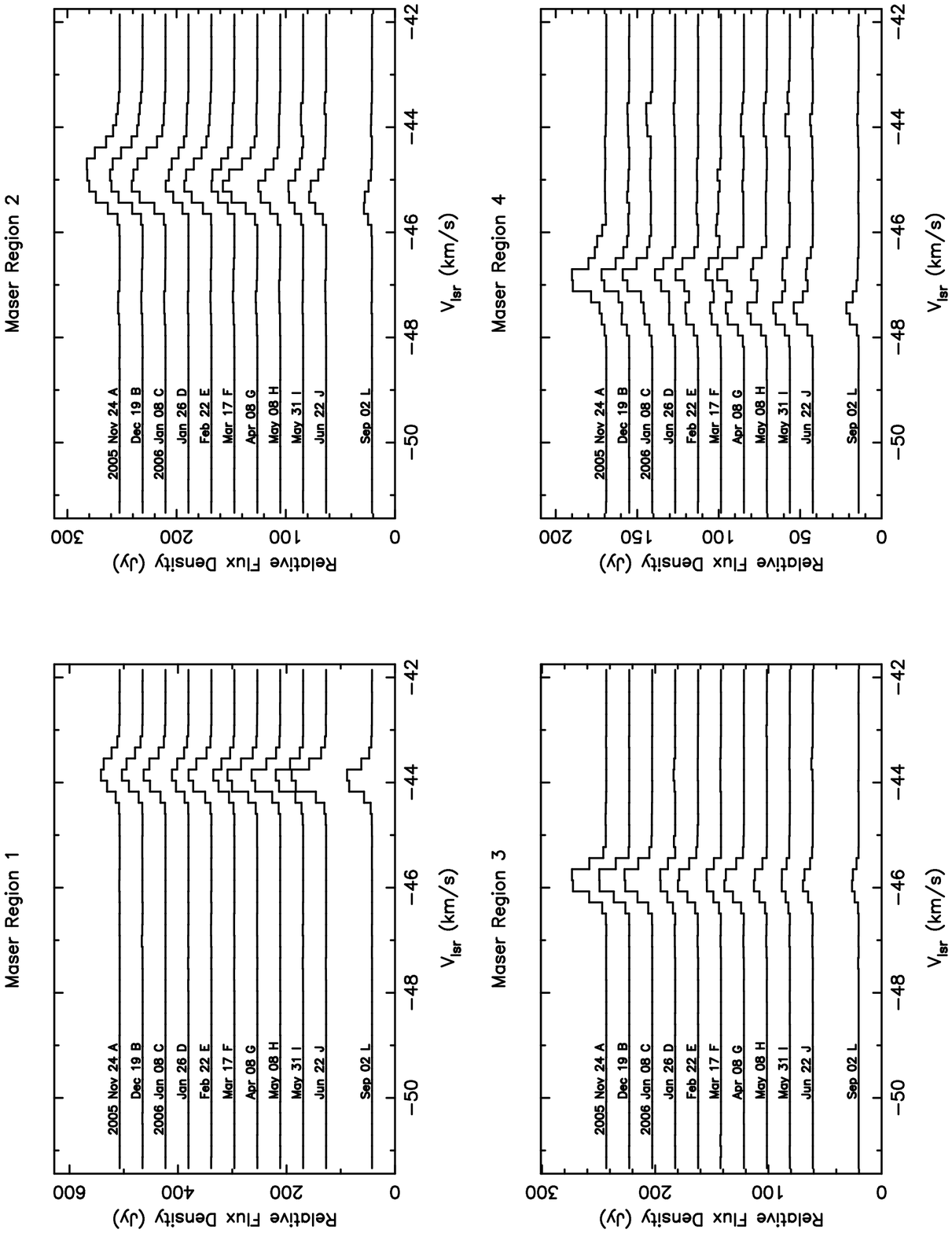}
\caption{Spectral distribution and temporal evolution of maser regions 1--4. \label{dynspecfig}}
\end{figure}

Figure \ref{integspecfig} shows the spectrum of \iras\, integrated
over all maser spot regions, for a representative set of epochs (A, D,
G, and J), indicating the overall variability of the source.  Figure
\ref{dynspecfig} shows separately the time-dependent spectra for maser
regions 1--4.  A variety of spectral structure and behavior is
observed.  Regions 1 and 3 show strong, stable, and simple spectral
structure. Region 2 shows a monotonic drift of five channels in the
peak of emission.  Region 4 shows two distinct spectral components:
the feature at $-45.0$ \kmps\ drifts by about one channel and fades,
while the feature at $-45.5$ \kmps, indistinct at epoch A, grows in
strength with epoch and is stationary in velocity.  The remaining
regions (not shown) are generally weaker, and show a similar range of
structure and variability.  Of these, maser region 5 is the most
stable. As shown below, the spectral structure and variability has
serious consequences for obtaining reliable parallax and proper motion
fits.

To obtain an estimate of the fundamental phase-referencing accuracy of
these observations, continuum images (bandwidth: 14 MHz) of the
secondary phase-reference calibrator, \prcalb (SNR$\sim$20), were
generated.  For the ensemble of epochs (excluding epoch K), the rms
peak position of \prcalb\ is $\pm$30 $\mu$as in both right ascension
and declination, with no evidence of any significant systematic
motion. (Without the macroscopic troposphere solution applied, the
declination rms increases by a factor of 2, indicating the importance
of this calibration.)  This estimate establishes a basis for
estimating the astrometric precision (per SNR) of the maser
observations.  Specifically, a SNR of $\sim$20 nominally corresponds
to an astrometric precision of $\sim$30 $\mu$as.  

This error estimate cannot be transferred and scaled directly to the
observations of \iras.  A variety of systematic errors render
themselves differently for \iras\ and \prcalb.  \prcalb\ is only
0.66$^\circ$ from \prcala, the primary calibrator, less than half the
1.41$^\circ$ \prcala--\iras\ separation, and so the phase-referencing
calibration is nominally more accurate for \prcalb.  Furthermore, near
transit at the center of the VLBA, the position angle of the
\prcala--\prcalb\ separation is mostly in azimuth while that of
\prcalb--\iras\ is mostly in zenith angle (see Figure \ref{calgeo}),
making any residuals in the zenith angle-dependent macroscopic
troposphere calibration more important for \iras.  As described below,
these and other systematic errors place more stringent limits on the
astrometric precision of \iras, despite the fact that most of the
maser spots used in the parallax fitting have higher SNR than \prcalb.

\section{The Fit for Annual Parallax and Proper Motion}

Parallax and proper motion calculations rely on measurements of the
time-dependent direction of consistently identified, small, physically 
coherent objects relative to a distant, presumed stationary,
background reference, in this case, the phase reference calibrator.
Masers are among the most compact observable objects available, but at
the resolution of the VLBA, they are generally not strictly
point-like.  Furthermore, it is not clear to what degree masers can be
characterized as distinct physical objects since they likely arise not
from fixed discrete objects, but rather from coherent velocity
structures within a presumably larger physical structure (e.g., gas
cloud).

Indeed, most of the maser spots in \iras\ are clearly not strictly
point-like and are variable in their marginally resolved structure and
flux density from epoch to epoch.  Furthermore, the spectral structure
of each maser spot is generally non-trivial, with the peak for some
spots clearly drifting substantially in velocity from epoch-to-epoch.
These factors make reliable epoch-to-epoch identification of distinct
physical features difficult.  Therefore, rather than fitting
complicated multiple point and resolved Gaussian model components to
the images or visibilities, and thereby introducing an additional
unphysical component in the analysis, we have adopted the simple peak
position (fitted from image pixels) as the best estimate of the maser
position, and have accepted the inevitable intrinsic variation of this
estimate (a form of traditional confusion within the synthesized beam)
as an additional source of error in our analysis.

For each of eleven distinct spots in 9 spot regions, and in each
channel with significant power (see below), we have fitted the
sequence of peak positions for parallax and two-dimensional proper
motion, according to the standard equation:
\begin{eqnarray}
\nonumber
\Delta\alpha\cos\delta & = & \Pi f_{\alpha} + \mu_{\alpha} t + \alpha_{o} \\
\Delta\delta & = & \Pi f_{\delta} + \mu_{\delta} t + \delta_{o} 
\end{eqnarray}
\noindent
where $\Pi$ is the annual parallax, $t$ is time,
$(\mu_{\alpha},\mu_{\delta})$ is the 2-dimensional proper motion,
$(\alpha_{o},\delta_{o})$ denote the maser position at $t=0$, and
$(f_{\alpha},f_{\delta})$ is a 2-dimensional function of direction and
time describing the parallax ellipse (the reflection of the Earth's
orbit onto the sky).  In using this equation, we nominally assume that
any non-linear internal motions with the maser spot regions are
negligible.

\begin{figure}
\includegraphics[angle=-90,scale=0.8]{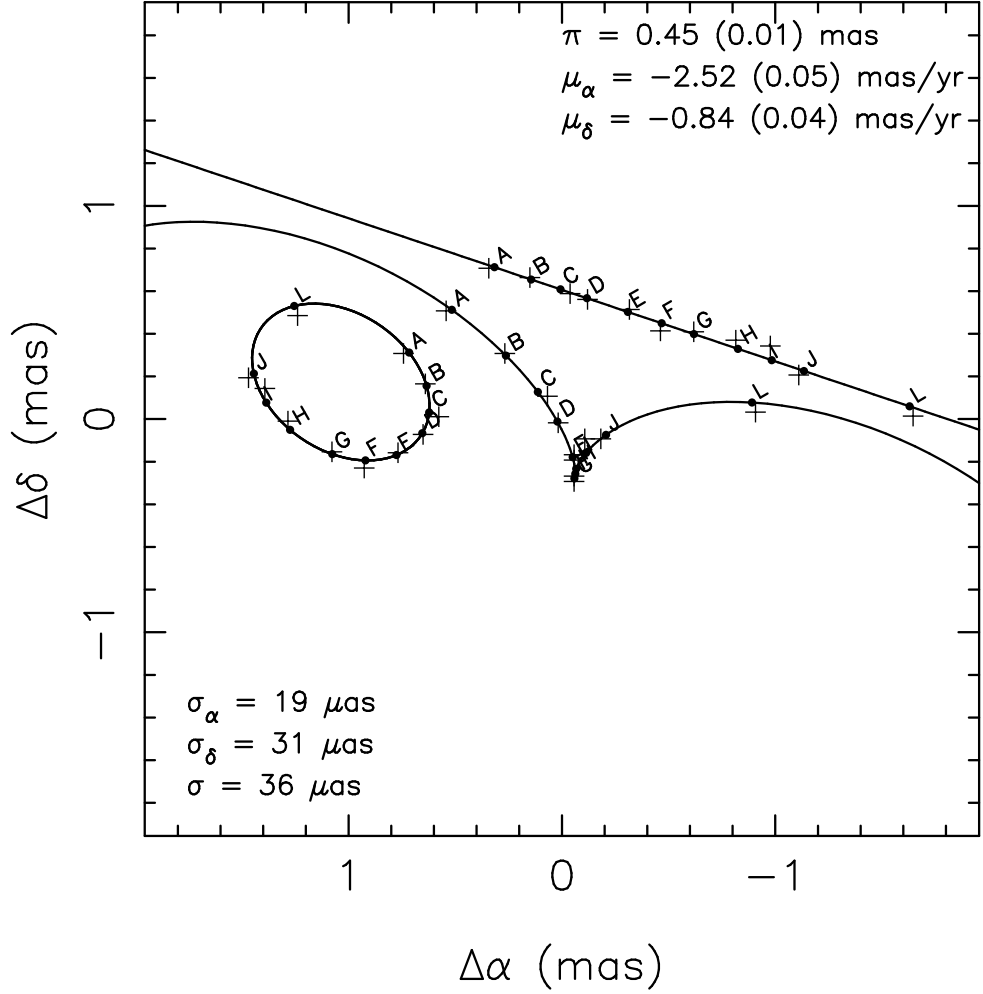}
\caption{The parallax fit for maser region 3 at $V_{LSR}=-46.0$ \kmps. 
The central curve shows the combined parallax and proper
motion evolution of the maser (crosses), with the fit overlaid (line
and filled dots). The top curve shows the proper motion with parallax
removed and the bottom curve (ellipse) shows the parallax with the
proper motion removed (same symbols).  \label{parallaxskyfig}}
\end{figure}

\begin{figure}
\includegraphics[angle=-90,scale=0.7]{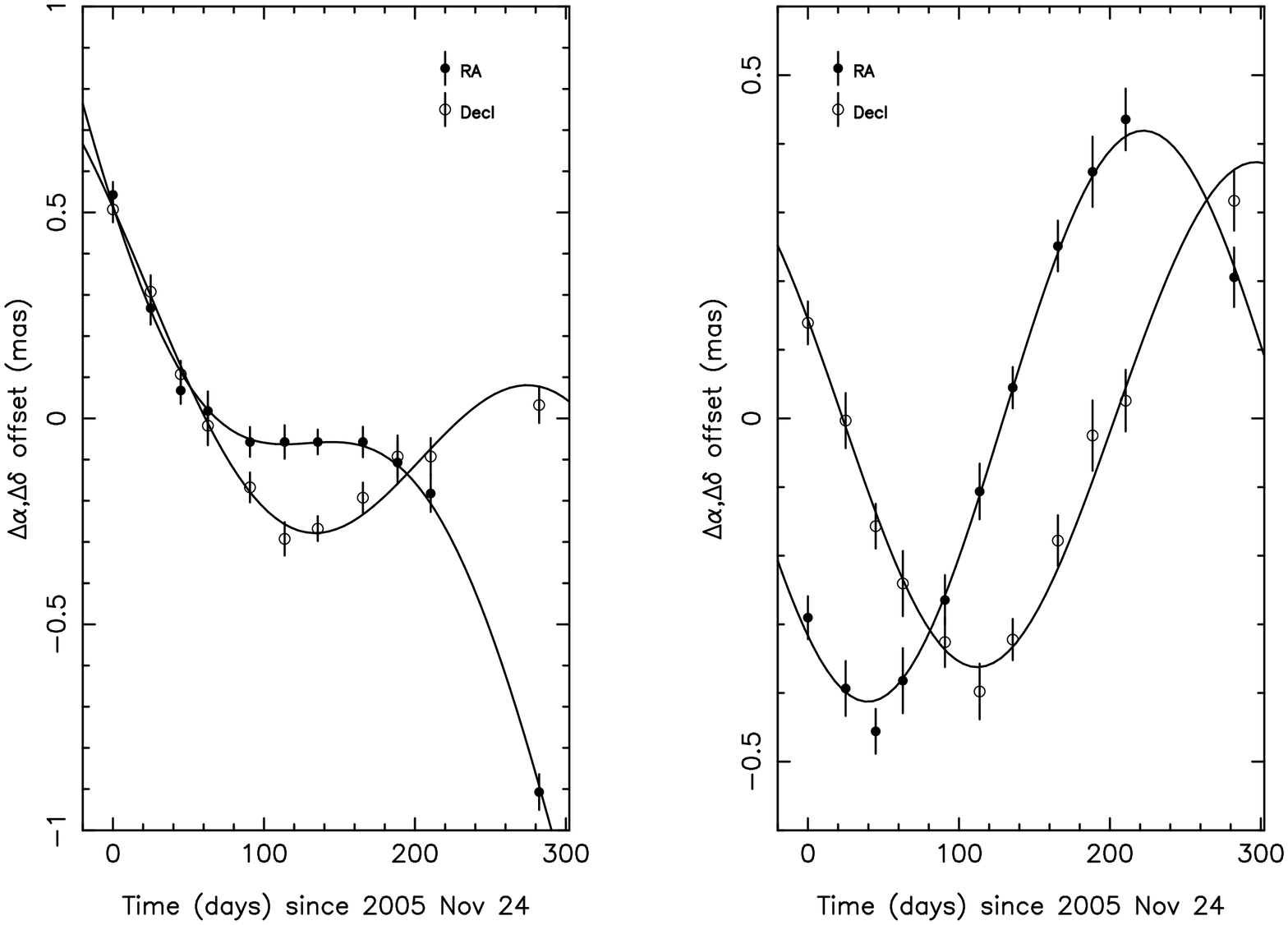}
\caption{Left: Positional evolution of maser region 3 at $V_{LSR}=-46.0$ \kmps\ 
in right ascension and declination.  Right: Positional
evolution with the proper motion removed. The error bars reflect the
relative weights used in the fitting, conservatively scaled so that
their minimum is 30 $\mu$as (see text).  \label{parallaxtimefig}}
\end{figure}

This yielded 44 estimates of the proper motion and parallax of
\iras, and the results are listed in Table \ref{tab3}.  An example fit 
is shown in Figures \ref{parallaxskyfig} and \ref{parallaxtimefig}.


\begin{deluxetable}{crrlrrrrrr}
\tabletypesize{\scriptsize}
\rotate
\tablecaption{Parallax Fits\label{tab3}}
\tablewidth{0pt}
\tablehead{
\colhead{} & \colhead{} & \colhead{} & \colhead{} & \colhead{} & \multicolumn{2}{c}{Proper Motion (Error)} & \multicolumn{3}{c}{Fit Residuals} \\
\colhead{Region} & \colhead{$V_{LSR}$\tablenotemark{a}} & \colhead{$N_{epochs}$} & \colhead{Epochs} & \colhead{Parallax\tablenotemark{b} (Error)} & \colhead{R.A.} & \colhead{Decl.} & \colhead{R.A.} & \colhead{Decl.} & \colhead{Both} \\
\colhead{} & \colhead{(\kmps)} & \colhead{} & \colhead{} & \colhead{(mas)} & \colhead{(mas year$^{-1}$)} & \colhead{(mas year$^{-1}$)} & \multicolumn{3}{c}{($\mu$as)}
}
\tablenotetext{a}{These velocities are in the traditional IAU local standard of rest used to set observing frequencies.}
\tablenotetext{b}{Values in {\bf boldface} have been used in the final parallax estimate (see text).} 
\tablecolumns{10}
\startdata

\sidehead{Good fits:}

  1 & -43.4 & 11 & ABCDEFGHIJ L & {\bf0.44} (0.02) & -2.06 (0.08) & -1.88 (0.05) &  32 &  45 &  55 \\
  1 & -43.6 & 11 & ABCDEFGHIJ L & {\bf0.44} (0.02) & -2.10 (0.07) & -1.95 (0.05) &  29 &  38 &  47 \\   
  1 & -43.9 & 11 & ABCDEFGHIJ L & {\bf0.44} (0.02) & -1.94 (0.06) & -1.72 (0.04) &  31 &  29 &  42 \\   
  1 & -44.1 & 11 & ABCDEFGHIJ L & {\bf0.48} (0.02) & -2.05 (0.07) & -1.63 (0.05) &  23 &  43 &  48 \\   
  1 & -44.3 & 11 & ABCDEFGHIJ L & {\bf0.49} (0.03) & -2.09 (0.09) & -1.65 (0.07) &  28 &  56 &  62 \\   
 & & & & & & & & & \\
  3 & -45.8 & 11 & ABCDEFGHIJ L & {\bf0.48} (0.01) & -2.55 (0.05) & -0.89 (0.04) &  19 &  30 &  36 \\   
  3 & -46.0 & 11 & ABCDEFGHIJ L & {\bf0.45} (0.01) & -2.52 (0.05) & -0.84 (0.04) &  19 &  31 &  36 \\   
  3 & -46.2 & 11 & ABCDEFGHIJ L & {\bf0.45} (0.01) & -2.52 (0.05) & -0.89 (0.04) &  18 &  32 &  37 \\   
 & & & & & & & & & \\
 4a & -46.6 &  6 & ABCDEF       & 0.50 (0.06) & -2.67 (0.13) & -1.92 (0.25) &  26 &  19 &  32 \\   
 4a & -46.8 &  6 & ABCDEF       & 0.49 (0.06) & -2.66 (0.13) & -1.97 (0.25) &  25 &  20 &  32 \\   
 4a & -47.2 &  6 & ABCDEF       & 0.50 (0.05) & -2.62 (0.10) & -1.93 (0.20) &  23 &  11 &  25 \\   
 & & & & & & & & & \\
 4b & -47.4 &  8 & \phm{ABC}DEFGHIJ L & {\bf0.43} (0.02) & -2.34 (0.08) & -1.85 (0.08) &  18 &  27 &  32 \\   
 4b & -47.6 &  8 & \phm{ABC}DEFGHIJ L & {\bf0.44} (0.02) & -2.40 (0.08) & -1.88 (0.08) &  22 &  23 &  32 \\   
 & & & & & & & & & \\
  5 & -43.2 & 10 & ABCDEFGHIJ   & {\bf0.50} (0.02) & -2.65 (0.09) & -1.05 (0.06) &  32 &  19 &  38 \\   
  5 & -43.4 & 10 & ABCDEFGHIJ   & {\bf0.47} (0.03) & -2.48 (0.13) & -1.18 (0.08) &  40 &  30 &  50 \\   
 & & & & & & & & & \\
\tablebreak

\sidehead{Good fits (cont):}
 7b & -43.9 &  7 & ABCDEFG      & 0.62 (0.05) & -5.79 (0.17) & -0.53 (0.19) &  25 &  28 &  38 \\   
 7b & -44.1 &  7 & ABCDEFG      & 0.51 (0.04) & -5.33 (0.12) & -0.77 (0.14) &  20 &  18 &  27 \\   
 7b & -44.3 &  7 & ABCDEFG      & 0.51 (0.04) & -5.29 (0.12) & -0.77 (0.13) &  15 &  19 &  25 \\   
 7b & -44.5 &  7 & ABCDEFG      & 0.50 (0.04) & -5.29 (0.14) & -0.81 (0.14) &  21 &  17 &  27 \\   
 & & & & & & & & & \\
  8 & -47.4 &  7 & ABCDEFG      & 0.59 (0.10) & -5.51 (0.31) & -1.23 (0.34) &  61 &  48 &  77 \\   
  8 & -47.6 &  7 & ABCDEFG      & 0.56 (0.09) & -5.71 (0.28) & -1.18 (0.33) &  63 &  36 &  73 \\   
  8 & -47.9 &  7 & ABCDEFG      & 0.48 (0.09) & -5.65 (0.27) & -1.39 (0.32) &  63 &  28 &  69 \\   
  8 & -48.1 &  7 & ABCDEFG      & 0.41 (0.08) & -5.57 (0.24) & -1.60 (0.28) &  54 &  23 &  59 \\   
 & & & & & & & & & \\
  9 & -49.3 &  8 & ABCDEFGH     & {\bf0.43} (0.04) & -5.01 (0.16) & -1.15 (0.13) &  35 &  30 &  46 \\   
  9 & -49.5 & 10 & ABCDEFGHIJ   & {\bf0.43} (0.03) & -4.99 (0.13) & -1.21 (0.09) &  32 &  41 &  52 \\   
  9 & -49.8 & 10 & ABCDEFGHIJ   & {\bf0.48} (0.03) & -5.17 (0.11) & -1.24 (0.08) &  36 &  26 &  44 \\   
  9 & -50.0 &  8 & ABCDEFGH     & {\bf0.48} (0.03) & -5.14 (0.10) & -1.42 (0.09) &  20 &  20 &  28 \\   
\tablebreak

\sidehead{Suspect fits:}

  2 & -44.5 & 10 & ABCDEFGHIJ  & 0.46 (0.04) & -1.21 (0.16) & -2.48 (0.11) &  32 &  56 &  65 \\   
  2 & -44.7 & 10 & ABCDEFGHIJ  & 0.52 (0.04) & -1.53 (0.14) & -2.42 (0.09) &  34 &  47 &  58 \\   
  2 & -44.9 & 10 & ABCDEFGHIJ  & 0.61 (0.04) & -1.79 (0.17) & -2.83 (0.10) &  36 &  57 &  67 \\   
  2 & -45.1 & 10 & ABCDEFGHIJ  & 0.48 (0.07) & -1.22 (0.28) & -3.50 (0.16) &  58 & 100 & 115 \\   
  2 & -45.3 & 10 & ABCDEFGHIJ  & 0.35 (0.10) & -0.72 (0.41) & -3.73 (0.24) & 117 & 120 & 167 \\   
  2 & -45.5 & 10 & ABCDEFGHIJ  & 0.25 (0.15) & -0.92 (0.59) & -2.83 (0.36) & 144 & 239 & 279 \\   
 & & & & & & & & & \\
  6 & -44.7 & 10 & ABCDEFGHIJ  & 0.83 (0.51) & -4.40 (2.10) &-11.01 (1.11) & 161 & 901 & 915 \\   
  6 & -44.9 & 10 & ABCDEFGHIJ  & 1.11 (0.44) & -5.34 (1.70) &-10.22 (0.93) & 251 & 779 & 819 \\   
  6 & -45.1 & 10 & ABCDEFGHIJ  & 0.90 (0.45) & -4.61 (1.71) &-10.27 (0.99) & 211 & 797 & 824 \\   
  6 & -45.3 & 10 & ABCDEFGHIJ  & 0.82 (0.42) & -4.28 (1.56) &-10.18 (0.95) & 199 & 796 & 820 \\   
 & & & & & & & & & \\
 7a & -43.2 &  7 & ABCDEFG     & 0.16 (1.60) &-13.68 (5.07) &  0.25 (5.43 )& 908 & 889 &1271 \\   
 7a & -43.4 &  7 & ABCDEFG     &-2.89 (2.30) &  2.45 (6.92) &-14.79 (8.06) &1307 &1227 &1793 \\   
 7a & -43.6 &  7 & ABCDEFG     &-0.43 (1.79) &-11.18 (5.47) & -3.06 (6.35) & 961 &1014 &1397 \\   
 7a & -43.9 &  7 & ABCDEFG     &-1.06 (1.89) & -8.59 (6.32) & -4.96 (6.32) & 873 &1017 &1340 \\   
\enddata    

\end{deluxetable}

For the fitting, only channels with maser spot detections greater than
$\sim$1--2 Jy (corresponding to SNR $\gtrsim$ 10, typically) were included.
Additionally, after a trial fit, partial epoch ranges for some spots
were discarded where it was evident that the assumption of a
consistent discrete physical component following the relatively simple
parallax and proper motion trajectory was violated.  Presumably, the
peak derived from our finite-resolution observations shifted abruptly
as a result of intrinsic structural changes.  Epoch K was excluded
entirely since the Mauna Kea station, which provides the longest
baselines, did not participate.

The fits for three maser regions (2, 6, \& 7a) were rejected outright
from further analysis on the basis of poor consistency among channels
or unphysical (negative) parallax estimates. Each of these regions show
substantial spectral complexity, including variation in the velocity
of peak emission, indicating that they are non-ideal test-particles
for parallax measurements.  No degree of judicious epoch selection
makes these maser regions recoverable for parallax estimation purposes.

Determining proper epoch- and maser-dependent weights for the fitting
process is non-trivial, since accurate {\em total} errors, including
systematics, for the peak position estimates are not readily available
{\em a priori}.  Formally, for a well-isolated point source with no
systematic errors, the precision of the peak position estimates would
be inversely proportional to the SNR of the peak emission, and the
resulting formal weights would be proportional to the {\em square} of
the SNR.  In practice, such an estimate of astrometric precision must
be a lower limit due to systematic errors arising from the
complicating maser structure considerations noted above, as well as to
per-epoch residuals in the macroscopic troposphere and
phase-referencing calibration.  Additionally, the wide range of flux
densities among maser regions and epochs ($\sim$2--60 Jy) result in a
range of weights of 2 orders of magnitude or more for some maser
spots.  Such weights could have a deleterious effect on the effective
sampling of the parallax ellipse during the year, and lead to
complicated biases in the parallax estimates.  Therefore, we adopted
weights proportional to SNR (i.e., the square root of the formal
weights) for the final fits.  For most of the fits, this choice was
not consequential; use of the formal weights yields parameter
estimates not significantly different from that obtained with SNR
weights.  Our parallax measurement is not limited by per-epoch SNR,
but rather by {\em per-epoch} systematic errors in the astrometry.

While the proper motions may be expected to vary among the spot
regions (and even among channels in a single region) due to intrinsic
internal motions, the parallax estimates should be consistent for all
of them since the range of maser emission is very small ($\lesssim300$
mas) relative to its distance (kiloparsecs), and is unlikely to be
significantly deeper than it is wide.  In practice, the variation of
the parallax estimates among maser spot regions and channels is more
complicated.  While several (but not all) maser regions show impressive
internal consistency, the dispersion of parallax estimates over the
entire ensemble is considerably larger than the formal errors would
imply.  This fact is related to residual systematic errors affecting
the fits which originate in the complexity and evolution of the
spectral and spatial structure of the maser regions, residual
calibration errors, and the details of the time-dependent sampling of
the parallax ellipse.

At any single epoch, calibration errors will affect different maser
spot regions uniformly, and so are systematic, i.e., relative position
measurements among masers at one epoch have no information about the
magnitude of such errors.  Over many epochs (up to eleven, in this
case), however, we assume such errors can be treated as approximately
stochastic.  Similarly, after excising obviously discrepant epochs as
described above, we treat any {\em non-linear} intrinsic variations in
the peak position due to internal structural and flux density
evolution as approximately stochastic over many epochs. These errors
limit the precision of the parallax estimate, but are unlikely to bias
it substantially.  With up to eleven epochs per fit, we assert that
the rms residuals determined in the fits are a reasonable estimate of
the mean per-epoch astrometric precision achieved in our observations,
typically $\sim20-30\mu$as in both dimensions.  The error bars in
figure \ref{parallaxtimefig} reflect the relative weights used in the
fit, and are conservatively scaled so that their minimum is 30 $\mu$as.

\begin{figure}
\includegraphics[angle=-90,scale=0.7]{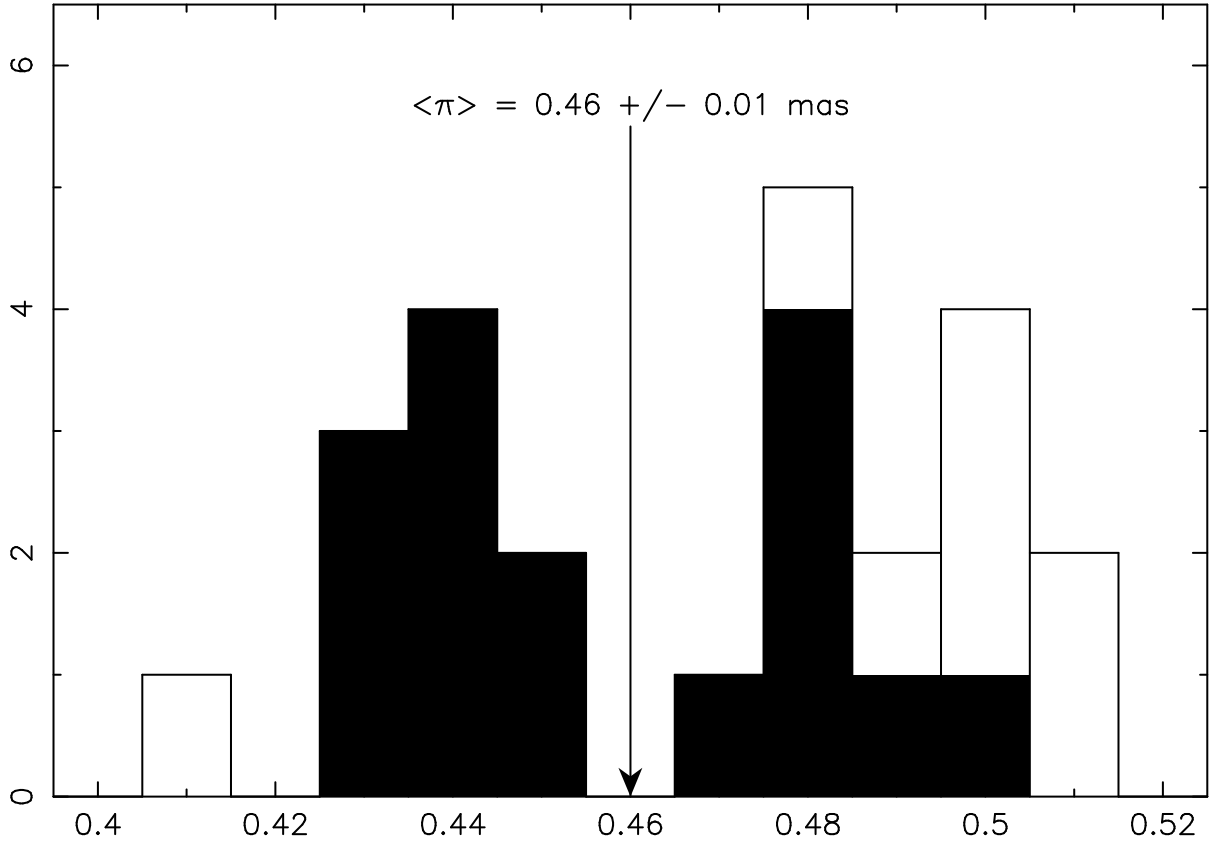}
\caption{The distribution of maser parallax estimates.  The open
histogram indicates estimates biased by poor sampling of the 
parallax ellipse (see text).  The filled histogram includes
only the estimates used to obtain the final parallax estimate of
0.46$\pm$0.01 mas. \label{parallaxhist}}
\end{figure}

Of the remaining 27 fits in 8 maser regions, most yield internally
consistent parallax and proper motions (with channel), but the overall
range of parallax results remains relatively large: 0.41-0.62 mas, with formal
errors in the range 0.01-0.1 mas (see Figure \ref{parallaxhist} which
shows the distribution in the range 0.40-0.52 mas).  In addition to
several obvious outliers (there are three such outliers greater than
0.55 mas and not shown in Figure \ref{parallaxhist}), there is some
evidence of maser region-dependent systematic errors.  A careful
analysis of the time-dependent weighted sampling shows that those
maser regions for which the later epochs were undetected or discarded
tend to yield systematically higher parallax estimates (near or in
excess of 0.50 mas) compared to those with more complete sampling.
Fitting trials on the completely sampled maser regions (e.g., region
3) using {\em only} the early epochs confirm this effect.  We
therefore have discarded maser regions 4a, 7b, and 8 as insufficiently
sampled and inadequate for further consideration.  The remaining 16
maser region channels are shown in the filled histogram of Figure
\ref{parallaxhist}.  The distribution still appears bimodal, but the
two states cannot be segregated according to maser region.  
The unweighted dispersion of the remaining measurements is 0.02 mas, 
broadly consistent with the typical formal errors in the individual fits,
which arise from a combination of pure statistics and per-epoch
systematics that are stochastic over many epochs.

\begin{figure}
\includegraphics[angle=-90,scale=0.7]{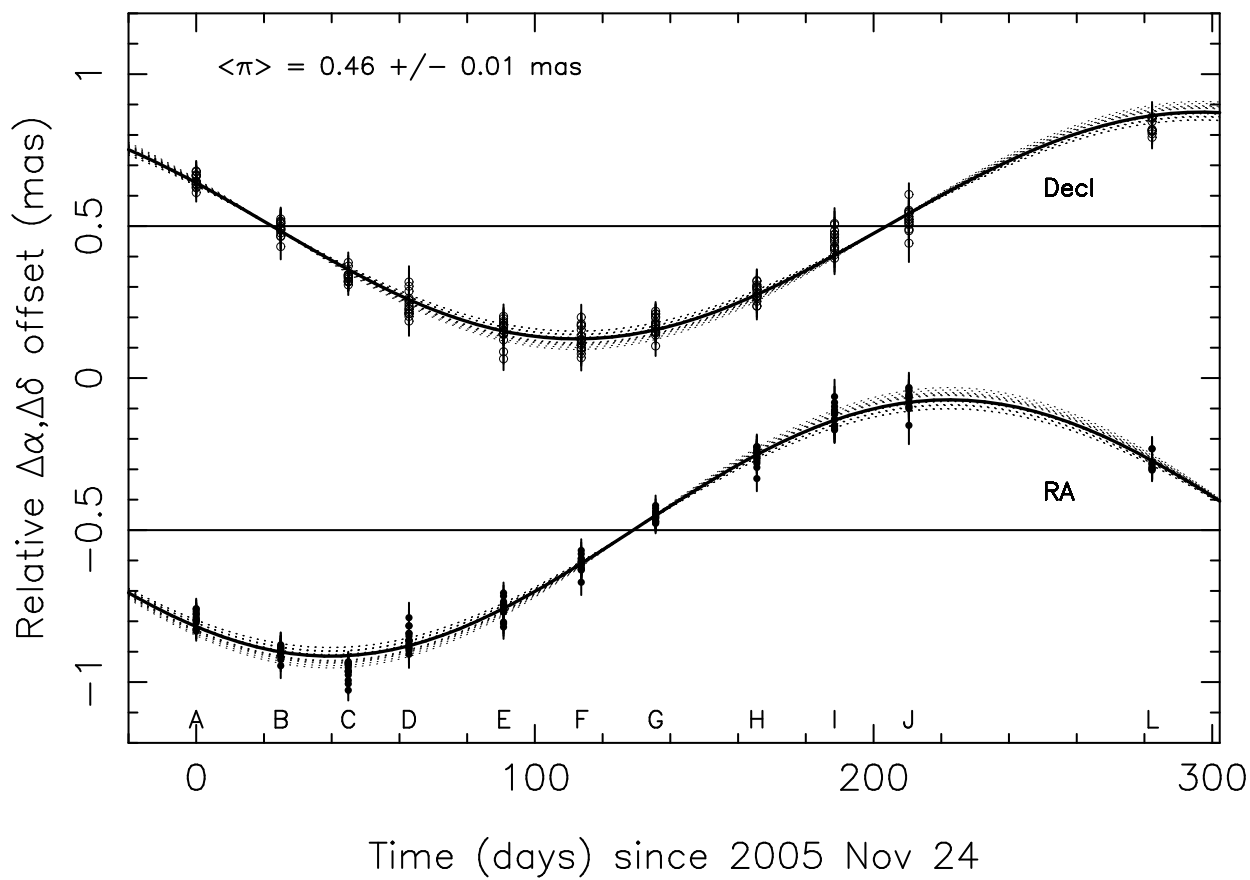}
\caption{The final parallax estimate's signature in right ascension and declination 
as a function of time (bold curve), with the per-channel observed
data (filled and open dots, with error bars as in Figure \ref{parallaxtimefig})
and the range of individual channel solutions (hashed region) overlaid (the
individual proper motions and relative position offsets have been removed).
The right ascension and declination data and curves have been displaced
from zero for clarity.  Epoch labels are listed along the bottom. \label{finalest}}
\end{figure}

The individual parallax solutions for the remaining 16 maser region
channels have been averaged with weights derived from their formal
errors to yield a net parallax estimate of 0.458$\pm$0.005 mas.
Figure \ref{finalest} shows the final parallax signature in right
ascension and declination as a function of time (bold curve) Also
plotted are the per-channel data (filled and open dots, with error
bars as in Figure \ref{parallaxtimefig}) and the range of individual
channel fits (hashed region); the fitted proper motions and relative
position offsets from the individual solutions have been removed.  It
is clear that several epochs (e.g., epoch C in right ascension, epochs
I and L in declination) are affected by a substantial residual
systematic offset in one dimension or the other, but that for most
epochs, the magnitude of these systematics is similar or smaller than
the statistical distribution of measurements. Since the parallax
solution has been derived directly from the data, it is difficult to
accurately ascertain the true balance between the per-epoch systematic
and statistical contributions.  Although the formal errors in each
individual maser channel solution nominally account for both sources
of error (assuming the per-epoch systematics are approximately
stochastic over many epochs), we nonetheless conservatively double the
formal error estimate and adopt a final parallax estimate of
0.46$\pm$0.01 mas.  The distance to \iras\ is therefore 2.17$\pm$0.05
kpc, a measurement with a precision of $<$3\%.

\section{Implications for Galactic rotation}
\label{GalRotAnalysis}

With direct measurements of the distance and three-dimensional space
velocity of objects such as \iras\ using the VLBA, it becomes possible
to more accurately evaluate the veracity of the kinematic distances
derived from Galactic rotation curves, and to do so over a larger
portion of the Galaxy than with any other available means.  This is
especially important outside the solar circle where rotation curve
analyses have had to rely principally on photometric and other
indirect distance estimates.

A large range of distance estimates for \iras\ have been used in the
literature for various astrophysical analyses.  Most tabulated
distances have been in the range $\sim4$--$6$ kpc, and are presumably
kinematic in origin.  Our \water\ maser parallax distance of 2.17 kpc
is, in fact, most consistent with the oldest published distance
estimate, a photometric distance of 1.7 kpc \citep{Neckel84}.

The direct geometric distance estimate reported here is more than a
factor of two closer than the kinematic distance of $\sim$4.7 kpc
implied by a radial velocity of $-46$ \kmps\ using the rotation curve
of \citet{Brand93} (henceforth, BB93).  However, those authors
acknowledge systematic offsets from their rotation curve in the sample
they use to derive it (HII regions and reflection nebulae), including
the fact that it overestimates the (galactocentric) orbital velocities
of most objects within $\sim$3 kpc outside the solar circle.
Furthermore, at the Galactic longitude and distance of \iras, in
particular, their line-of-sight velocity residuals show a systematic
deficit of $\sim$25
\kmps;  their rotation curve nominally predicts a radial velocity of
$-$25 \kmps\ for \iras, $\sim$21 \kmps\ slower than observed.  Similar
large-scale systematic residuals occur at many outer Galaxy
longitudes, and they attribute this to non-circular Galactic
orbits, or more generally, the inadequacy of an axisymmetric model for
the velocity field of the outer Galaxy.  In short, although the
rotation curve predicts Galactic orbital velocities with errors of
only 10$-$15\% (out of 200$-$250 \kmps), it is a much poorer predictor
of {\em line-of-sight} velocities measured at the Sun, and so
kinematic distance estimates for the outer Galaxy will be generally
unreliable.

\begin{figure}
\includegraphics[angle=-90,scale=0.7]{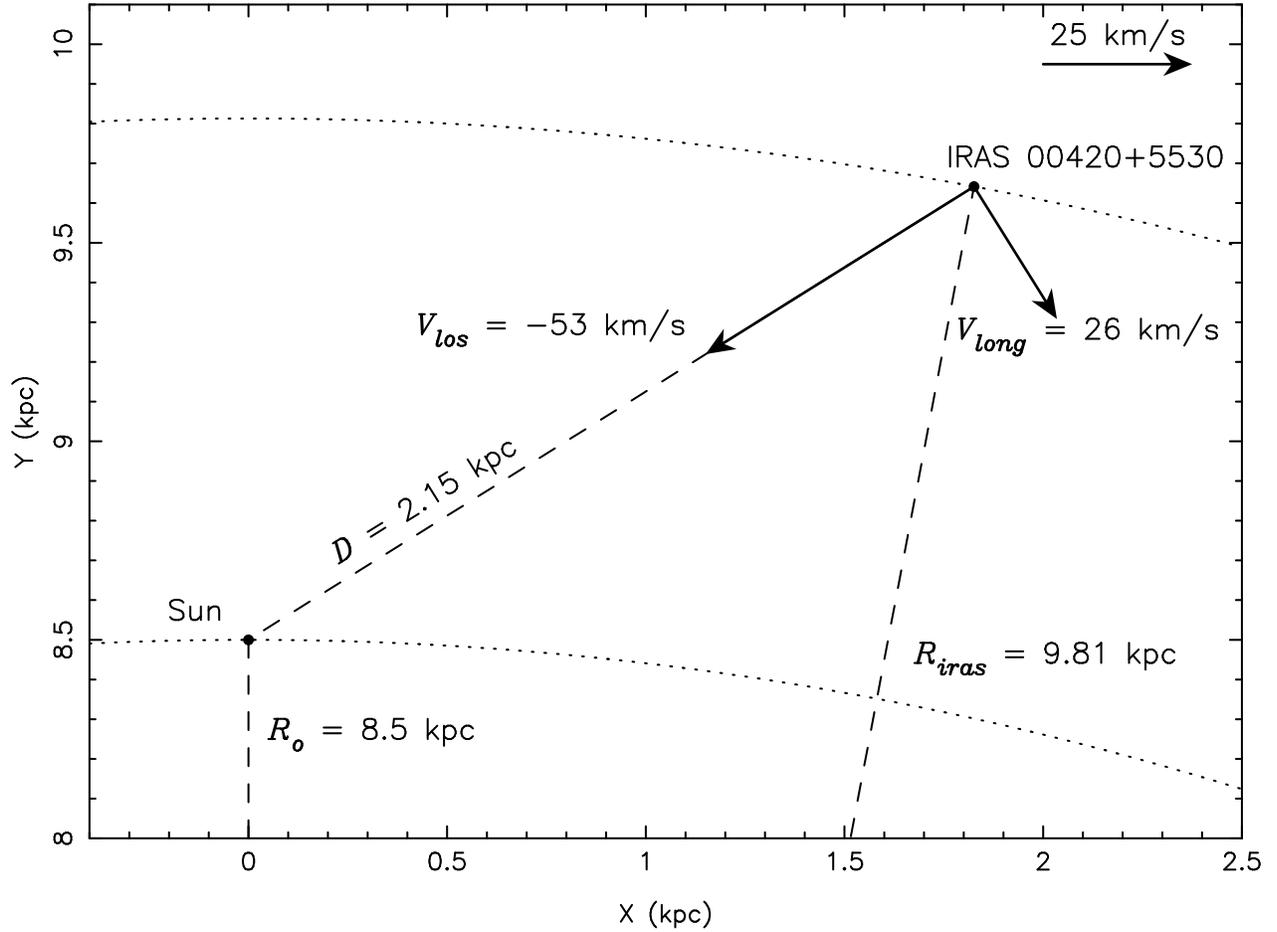}
\caption{Plan view of the Galaxy (from above), showing
the observed relative positions and {\em heliocentric} velocities of
\iras\ and the Sun.  All quantities are projected onto the Galactic 
plane in a heliocentric frame (the Sun is stationary).  The dotted lines
indicate the nominal circular galactocentric orbits of the Sun and \iras.
The dashed lines connect the Sun, \iras, and the Galactic center, which
lies at (0,0) kpc. \label{galplanobs}}
\end{figure}

Since our VLBA measurements provide the three-dimensional space
velocity (relative to the Sun) of \iras, it is possible to explore the
nature of its Galactic orbit in detail.  Adopting the maser component 
in region 3, $V_{LSR}=46.0$ \kmps\ as representative for \iras\, we find a {\em
heliocentric} line-of-sight velocity of $-52.5$ \kmps, and, at a
distance of 2.17 kpc, a (heliocentric) physical velocity of $-26.2$
\kmps in Galactic longitude and $-7.9$ \kmps\ in Galactic latitude (at the
distance of \iras, 1 milliarcsecond year$^{-1}$ $=$ 10.3 \kmps).
Figure \ref{galplanobs} illustrates the in-plane components of these
measurements in a plan view of the region of the Galaxy containing the
Sun and \iras.  Formally, the error in these velocities is $\sim$0.5
\kmps, but there is a full range of $\sim$10 \kmps\ in the
line-of-sight and latitude components, and somewhat less in the
longitude components among the brightest maser spot regions.  We assume
tentatively that the internal motion of this component is not large enough to 
excessively bias our analysis (this point will be addressed in greater
detail below).

The following analysis is limited by the accuracy of knowledge of the
absolute orbital speed of the Galaxy and the Sun's peculiar motion
relative to the LSR. We have provisionally assumed standard IAU
values for the LSR parameters, namely that it is $\Ro=8.5$ kpc
from the Galactic center and moving at $\To=220$ \kmps\ on a circular
orbit \citep{Kerr86}.  The Sun's peculiar motion
relative to the LSR has been assumed to be $U=10.0$ \kmps\ (toward the
Galactic center), $V=5.25$ \kmps\ (along the orbit), and $W=7.17$
\kmps\ (out of the plane) as measured by \citet{Dehnen98} using Hipparcos
data.  The calculations have be done in three dimensions, but with no
correction for the Sun's unknown distance from the Galactic plane.

\begin{figure}
\includegraphics[angle=-90,scale=0.7]{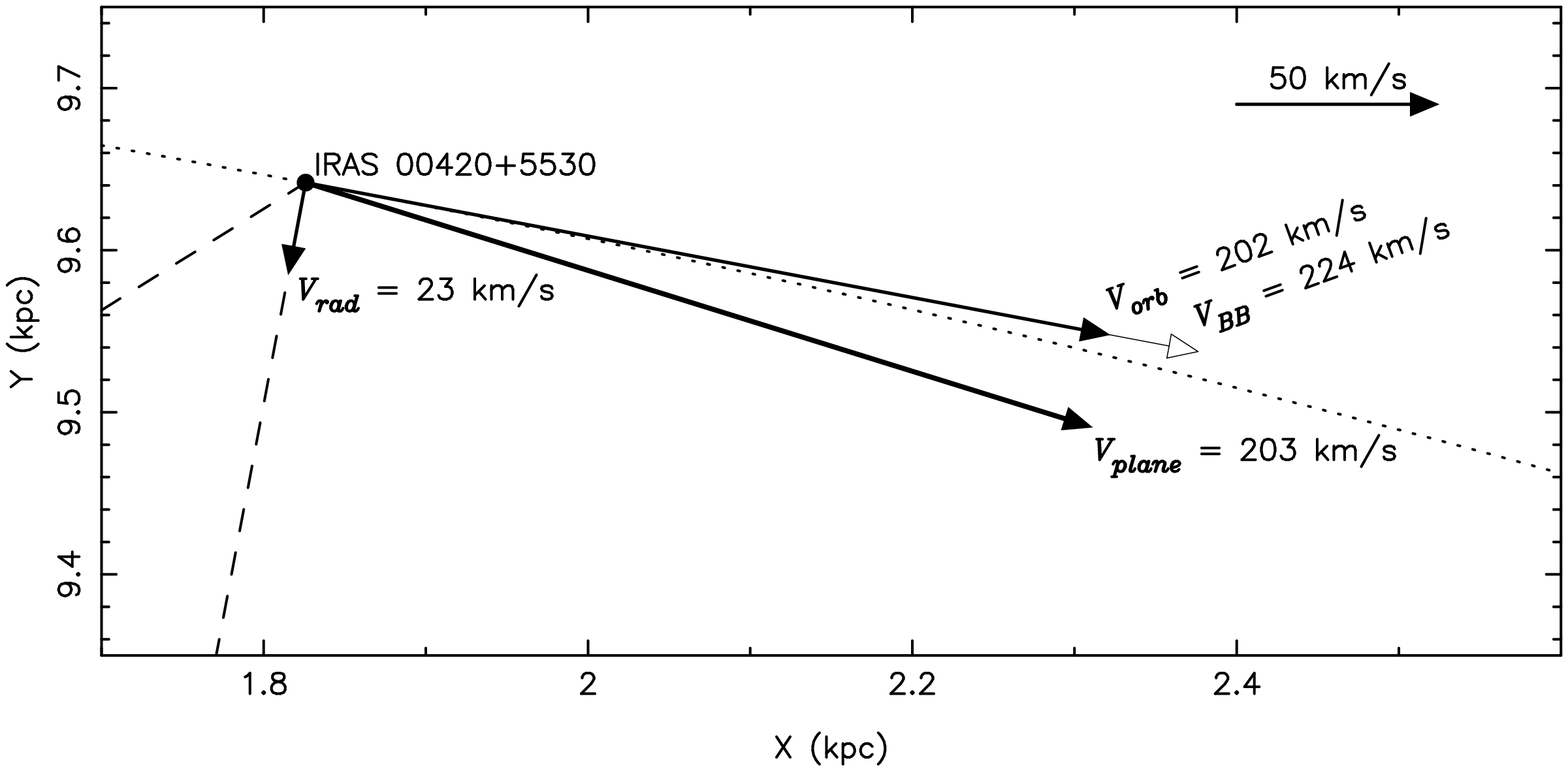}
\caption{Plan view showing the motion of \iras\ in a {\em galactocentric}
frame, having corrected for solar peculiar motion ($U=10$ \kmps,
$V=5.25$ \kmps, \& $W=7.17$ \kmps; see text) and assuming a circular
orbital velocity of 220 \kmps\ for the (solar) LSR.  The dotted line
indicates the nominally circular orbit at the galactocentric radius of
\iras.  The \citet{Brand93} rotation curve predicts an orbital speed
of 224 \kmps\ on this line.  The dashed lines connect \iras\ to the
Sun at (0,8.5) kpc and the Galactic center at (0,0) kpc. \label{irasgal}}
\end{figure}

Converting to the galactocentric frame by correcting for $\To$ and
solar peculiar motions, we find that \iras\ is moving toward the
Galactic center at 23.0 \kmps, around the Galaxy (circular component)
at 202.0 \kmps, and toward the plane (from below) at 5.8 \kmps.  The
in-plane components of \iras's velocity vector are shown in Figure
\ref{irasgal}. Thus, \iras\ is moving significantly slower than the
circular speed of $\sim$224 \kmps\ predicted by the BB93 rotation
curve, and has a substantial non-circular component toward the
Galactic center.

\begin{figure}
\includegraphics[angle=-90,scale=0.7]{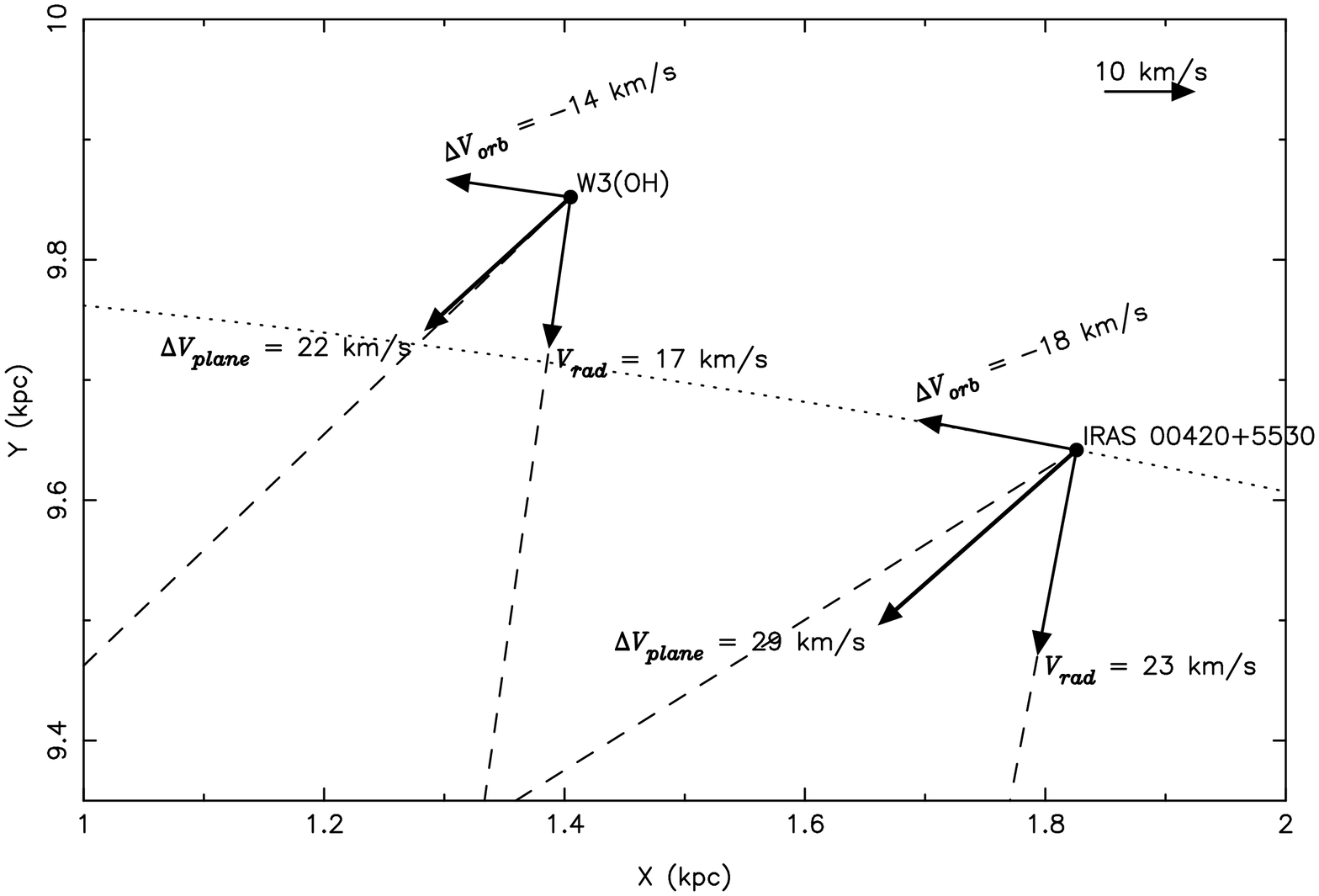}
\caption{Plan view showing the motion of \iras\ and W3(OH) \citep{Xu06} 
in a frame co-rotating with the Galaxy at 220 \kmps.  The dotted line
indicates the nominally circular orbit at the galactocentric radius of
\iras.  The dashed lines connect \iras\ and W3(OH) to the Sun at
(0,8.5) kpc and the Galactic center at (0,0)
kpc. \label{irascorot220}}
\end{figure}

By assuming a nominal orbital speed for \iras, we can shift our
velocity measurements to a local standard of rest orbiting the Galaxy
at the position of \iras.  Figure
\ref{irascorot220} shows the resulting in-plane peculiar motion of \iras,
assuming its orbital speed is 220 \kmps. (The BB93 rotation curve
prediction is insignificantly larger; essentially we nominally assume
the rotation curve is flat in this region of the Galaxy.)  Also shown
is the peculiar motion of W3(OH) (also in its {\em own} standard of rest) as reported by
\citet{Xu06}, also assuming an orbital speed of 220 \kmps.
The peculiar motions of \iras\ and W3(OH) are in almost exactly the
same direction, and very nearly in the direction of the Sun.  Combined
with the large-scale systematic line-of-sight velocity deficit
observed in the BB93 rotation curve residuals, this is highly
suggestive of a large-scale anomalous motion of the material in this
region of the Galaxy, i.e., the Perseus spiral arm, in a direction
toward the Sun.  Such anomalous motion could be related to the
formation and propagation of the Perseus spiral arm, and is evidence
of the inadequacy of axisymmetric models of Galactic rotation.

An obvious alternative to this conclusion is that the Sun has a large
velocity component in the direction of these objects, either in terms
of the LSR orbital speed, the solar peculiar motion, or both.  Indeed,
the orbital speed component of the Sun's velocity is the least
well-understood component of its motion due to uncertainties in the
overall scale of the Galaxy ($\Ro$ and $\To$), and the component of
solar peculiar motion in the direction of Galactic rotation.  However,
the velocity adjustments required of the alternative scenarios are
generally implausible.  Increasing the Galactic orbital speed for the
solar LSR to 250 \kmps\ (the value favored by \citet{Reid04} for $\Ro
= 8.5$ kpc according to measurements of the proper motion of Sgr
A$^*$) and maintaining the flat rotation curve out to \iras\ and
W3(OH) reduces the radial (toward the Galactic center) component of
the anomalous velocity by only $\sim$6 \kmps, without substantially
changing the orbital component.  The anomaly is even less sensitive to
adjusting {\em both} the orbital speed and $\Ro$ in such a way as to
keep the {\em angular} speed ($\To/\Ro$) of the solar LSR constant
\citep{Reid04}.  This insensitivity is a result of the relatively
small angle ($\sim$11$^\circ$) subtended at the Galactic center by the
Sun and \iras, and that this angle changes very slowly for a fixed
distance to \iras\ (our measurement) over the plausible range of $\Ro$
(8 $\pm$ 1 kpc).  Introducing a difference in nominal orbital
speed between the Sun and these sources sufficient to account for the
anomaly is also implausible, implying zero additional mass outside the
solar circle, and would not compensate for the radial component of the
anomaly.  Finally, a sufficient correction to just the solar peculiar
motion would require a change to the radial component of $\sim$20
\kmps, entirely inconsistent with the Hipparcos measurements of
\citet{Dehnen98}.

It is also possible that we have failed to model a large internal
motion of the \water\ masers in \iras\ that happens to lie in the
direction of the apparent anomalous motion, a possibility noted by
\citet{Hachisuka06} for W3(OH).  However, we consider this unlikely
given the relatively modest range of three-dimensional velocities
observed for the main \water\ masers in \iras, and that they are near
($\lesssim$ 5 \kmps) the systemic velocity indicated by other
molecular species in \iras.  The maser channel chosen for this
analysis (region 3, $-46.0$ \kmps) is reasonably representative of the
available information for \iras.  Other choices are possible, but
either can be excluded on grounds of insufficient constraints or do
not substantively change the result.  The maser regions excluded from
the parallax analysis have proper motion estimates biased toward more
negative velocities in longitude due to poor sampling of the parallax
ellipse in the late epochs.  Of the remaining 16 well-sampled maser
channels used in the parallax analysis, only region 9 differs in the
longitude component of its proper motion by more than $\sim$5
\kmps\ (at the longitude of \iras, equatorial and Galactic coordinates
are approximately parallel, so $\mu_{long}\simeq\mu_{ra}$).  The
dispersion in the latitude component is somewhat larger, but it is
irrelevant to the space velocity analysis within the Galactic plane.
Region 9 has a velocity in longitude $\sim$25 \kmps\ more negative than
region 3.  Alternatively choosing region 9 as representative ameliorates
the anomaly in the circular orbital component of the velocity found in
the above analysis for region 3 quite well, but also {\em increases}
the anomalous component in the direction of the Galactic center.
While we consider the remarkable consistency with the W3(OH) anomalous
space velocity compelling, it is not, by itself, sufficient evidence
of any interpretation of possible internal motions in \iras.
Without an absolute reference within the source, internal motion
biases cannot be ruled out, but an anomalous motion in the general
direction of the Sun and the Galactic center cannot be excluded by the
available \water\ maser proper motion data.  Distance and three-dimensional 
space velocity measurements of a larger sample of such objects in this
and other regions of the Galaxy \citep[e.g.,][]{Reid08} are necessary
to resolve such issues conclusively, and establish the true bulk motion
of material in the Perseus spiral arm.  Such observations could possibly
also better constrain the Galactic orbital speed and solar peculiar
motion, but will be limited by the isotropy and homogeneity (i.e., the
apparent lack of Galactic axisymmetry) of the available sample.

\citet{Sato08} have measured the distance to NGC~281---a star-forming
region only $\sim1^\circ$ from \iras---using VERA \citep{Honma00}
observations of
\water\ masers and a similar analysis (using the same phase-reference
calibrators).  They find a distance for NGC~281 of 2.82$\pm$0.24 kpc,
650 pc further than \iras, and postulate that these sources lie on
opposite sides of the surface of an HI super-bubble that is expanding
out of the plane of the Galaxy.  Their model is geometrically
appealing, and consistent with a substantial anomalous velocity
component toward the Sun for
\iras, on the near side of the bubble.  NGC~281 is on the opposite side
of the bubble, with line-of-sight velocity consistent with their model.
However, their model also predicts a large velocity out of the plane
for \iras, which we do not observe.  Unmodeled internal motions of
the \water\ masers in \iras\ perpendicular to the Galactic plane could
explain this discrepancy, but the available information cannot resolve
this ambiguity.  W3(OH) is not associated with the super-bubble in
their model.

\section{Conclusions}

Using the VLBA, we have measured the parallax and three-dimensional space
velocity of \water\ masers in \iras.  The parallax of 0.46$\pm$0.01 mas
provides a purely geometric distance estimate of 2.17$\pm$0.05 kpc
($<3$\%), a factor of two closer than popularly assumed kinematic
distance estimates suggest, yet consistent with an older, far-less accurate
photometric distance.  At this distance, the space velocity of
\iras\ shows a substantial anomalous component in the direction of the
Sun, consistent with similar observations of W3(OH), implying a
significant non-circular systematic motion of the Perseus spiral arm
of the Galaxy.

Parallax observations of larger samples of maser sources using the
VLBA have the potential to extend the distance estimates of a few \%
accuracy to a large fraction of the Galaxy, and will contribute to
more accurate non-axisymmetric models of Galactic rotation.

\acknowledgements
\noindent
The National Radio Astronomy Observatory is operated by Associated
Universities Inc., under cooperative agreement with the National
Science Foundation.  The authors wish to thank M.~Reid for useful
discussions.

\noindent
{\em Facilities:} \facility{VLBA}


\begin{thebibliography}{}
\bibitem[Brand \& Blitz(1993)]{Brand93}
    Brand, J. and Blitz, L. 1993, \aap\ 275, 67
\bibitem[Brand \etal,(1994)]{Brand94}
    Brand, J., \etal\ 1994, \aap\ Suppl 103, 541
\bibitem[Brisken \etal,(2002)]{Brisken02}
    Brisken, W.~F., Benson, J.~M., Goss, W.~M., \& Thorsett, S.~E. 2002, \apj\ 571, 906
\bibitem[Chatterjee \etal,(2004)]{Chatterjee04}
    Chatterjee S., Cordes, J.~M., Vlemmings, W.~H.~T., Arzoumanian, Z., Goss, W.~M., \& Lazio, T.~J.~W. 2004, \apj\ 604, 339
\bibitem[Dehnen \& Binney(1998)]{Dehnen98}
    Dehnen, W. \& Binney, J.~J. 1998, \mnras, 298, 387
\bibitem[Greisen(2003)]{Greisen03} 
    Greisen, E.~W.\ 2003, Astrophysics and Space Science Library, 285, 109
\bibitem[Hachisuka \etal,(2006)]{Hachisuka06}
    Hachisuka, K., \etal\ 2006, \apj\ 645, 337
\bibitem[Honma \etal,(2000)]{Honma00}
    Honma, M., Kawaguchi, N., \& Sasao, T. 2000, \procspie\ 4015, 624
\bibitem[Kerr \& Lynden-Bell(1986)]{Kerr86}
    Kerr, F.~J. \& Lynden-Bell, D. 1986, \mnras, 221, 1023
\bibitem[Ma \etal,(1998)]{Ma98}
    Ma, C., \etal\ 1998, \aj, 116, 516
\bibitem[Molinari \etal,(1996)]{Molinari96}
    Molinari, S., Brand, J., Cesaroni, R., \& Palla, F. 1996 \aap, 308, 573
\bibitem[Molinari \etal,(2002)]{Molinari02}
    Molinari, S., Testi, L., Rodr\'{i}guez, L.~F., \& Zhang, Q. 2002, \apj\ 570, 758
\bibitem[Neckel \& Staude(1984)]{Neckel84}
    Neckel, T. \& Staude, H.~J. 1984, \aap\ 131, 200
\bibitem[Reid (2008)]{Reid08}
    Reid, M.~J. 2008, \baas, 212, 6303
\bibitem[Reid \& Brunthaler(2004)]{Reid04}
    Reid, M.~J. \& Brunthaler, A. 2004, \apj, 616, 872
\bibitem[Reid \etal,(1999)]{Reid99}
    Reid, M.~J., Readhead, A.~C.~S., Vermeulen, R.~C., \& Treuhaft, R.~N 1999, \apj\ 524, 816
\bibitem[Sato \etal,(2008)]{Sato08}
    Sato, M., \etal\ 2008, \pasj, in press
\bibitem[Shepherd(1997)]{Shepherd97} 
    Shepherd, M.~C.\ 1997, Astronomical Data Analysis Software and Systems VI, 125, 77
\bibitem[van Langevelde \etal,(2000)]{vLangevelde00}
    van Langevelde, H.~J., Vlemmings, W., Diamond, P.~J., Baudry, A., \& Beasley, A.~J. 2000, \aap\ 357, 945
\bibitem[Xu \etal,(2006)]{Xu06} 
    Xu, Y., Reid, M.~J., Zheng, X.~W., \& Menten, K.~M. 2006, Science 311, 54
\bibitem[Zhang \etal,(2005)]{Zhang05}
    Zhang, Q., Hunter, T.~R., Brand, J., Sridharan, T.~K., Cesaroni, R., Molinari, S., Wang, J., \& Kramer, M. 2005, \apj\ 625, 864

\end{thebibliography}
\end{document}